\documentclass[aps,prd,showpacs,showkeys,reprint
]{revtex4-1} 
\usepackage{graphicx,dcolumn,bbm,amssymb,amsmath,amsfonts,hyperref,tabularx}
\usepackage{color}
\newcommand{\ud}{\mathrm{d}}

\newcommand{\om}{\omega_{\textrm{\tiny BD}}}

\begin{document}

\title{Dynamics and cosmological constraints on Brans-Dicke cosmology}

\author{Orest Hrycyna}
\email{orest.hrycyna@fuw.edu.pl}
\affiliation{Theoretical Physics Division, National Centre for Nuclear
Research,  Ho{\.z}a 69, 00-681 Warszawa, Poland}
\author{Marek Szyd{\l}owski}
\email{marek.szydlowski@uj.edu.pl}
\affiliation{Astronomical Observatory, Jagiellonian University, Orla 171,
30-244 Krak{\'o}w, Poland}
\affiliation{Mark Kac Complex Systems Research Centre, Jagiellonian
University, Reymonta 4, 30-059 Krak{\'o}w, Poland}
\author{Micha{\l} Kamionka}
\email{kamionka@astro.uni.wroc.pl}
\affiliation{Astronomical Institute, University of Wroc{\l}aw,
Kopernika 11, 51-622 Wroc{\l}aw, Poland}

\date{\today}

\begin{abstract}
We investigate observational constraints on the Brans-Dicke cosmological model using observational data coming from distant supernovae type Ia, the Hubble function $H(z)$ measurements, information coming from the Alcock-Paczy{\'n}ski test, and baryon acoustic oscillations. Our analysis is based on the modified Friedmann function resulting form dynamical investigations of Brans-Dicke cosmology in the vicinity of a de Sitter state. The qualitative theory of dynamical systems enables us to obtain three different behaviors in the vicinity of this state.  We find for a linear approach to the de Sitter state $\om=-0.8606^{+0.8281}_{-0.1341}$, for an oscillatory approach to the de Sitter state  $\om=-1.1103^{+0.1872}_{-0.1729}$, and for the transient de Sitter state represented by a saddle-type critical point $\om=-2.3837^{+0.4588}_{-4.5459}$. We obtain the mass of the Brans-Dicke scalar field at the present epoch as $m_{\phi}\sim H_{0}$. The Bayesian methods of model comparison are used to discriminate between obtained models. We show that observational data point toward vales of the $\om$ parameter close to the value suggested by the low-energy limit of the bosonic string theory.
\end{abstract}

\pacs{04.50.Kd, 98.80.-k, 95.36.+x, 95.35.+d}
\keywords{modified theories of gravity, cosmology, dark energy, dark matter}

\maketitle

\section{Introduction}

Composing the standard cosmological model ($\Lambda$CDM model) we assume that the general relativity describes universe and we postulate validity of the cosmological principle. This model is the best description of the current universe as indicated by implementing the Bayesian methods of model selection to simple theoretical models \cite{Szydlowski:2006ay}. Attempts to explain the present universe in terms of the standard cosmological model is justified by a pragmatic approach of a simple two parameter model. Such a model corresponds to what in physics we know as effective theories, like a standard model in particle physics. In this model as well as in $\Lambda$CDM model there are parameters which value should be obtained from a more fundamental theory or determined by observations. In cosmology the role of such parameters play the density parameters. Unfortunately, the nature of some parameters describing the dark side of the Universe (dark energy and dark matter) is unknown. From our point of view it means that in the construction of the standard cosmological model the cosmological constant term plays only the role of a useful fiction; i.e., the $\Lambda$CDM model describes cosmological observations well but unveils nothing about the nature of the cosmological constant. Adopting the methodology of an effective theory may shed some light on the nature of parameters revealing hints toward a more fundamental theory which we are looking for.

Because of well-known problems with the cosmological term in the standard cosmological model related with its substantial interpretation we are looking for a solution of the conundrum of acceleration of the current Universe in the framework of Brans-Dicke theory of gravity \cite{Brans:1961sx} (see also \cite{Clifton:2011jh}).
In this theory a gravitational interaction is described in terms of both a scalar field and the metric. The scalar field plays an important role in description of the early universe (inflation) as well as the late time cosmic evolution (quintessence). Moreover following recent Planck observations it is found that the time-varying equation of state with the constant additive contribution is favored when the astrophysical data are taken into account \cite{Ade:2013zuv}. 

In this framework the difficulty is to obtain a time-varying form of equation of state. In all these applications the scalar field is treated as a source. Usually it is assumed that they are not free and interact with itself via some potential function. Then in scalar field cosmology the main problem is to determine the unknown form of this potential. This problem is passed by assuming different, chosen {\it a priori} forms of the potential function of the scalar field.

In the Brans-Dicke theory, which is a scalar-tensor theory of gravity, a scalar field does not play the role of a substance but is rather a integral part of the gravitational sector. In this description a free parameter $\om$ appears as a consequence of the effective theory approach.

The value of this parameter can be constrained using the astronomical observations and astrophysical experiments. In the scale of the solar system the Cassini spacecraft mission experiment gave a very stringent bound on $\om > 4000$ for spherically symmetric solutions in the parametrized post-Newtonian (PPN) formalism  \cite{Will:book,Will:2005va,Bertotti:2003rm}. On the other hand the data from the cosmological experiments conducted during the WMAP and Planck missions gave substantially lower values of limits on the parameter $\om$.  Liddle et al. in \cite{Liddle:1998ij} studied the transition form radiation domination to matter domination epoch in Brans-Dicke theory and showed how the Hubble length at equality depends on the coupling parameter $\om$ for large values of this parameter. Acquaviva et al. in \cite{Acquaviva:2004ti} using structure formation constraints found lower bound $\om>120$ at $95\%$ confidence level. Recently Avilez and Scordis, using CMB data, have obtained the smaller value of the limit $\om > 692$ at a $99\%$ confidence level \cite{Avilez:2013dxa}. Li et al. \cite{Li:2013nwa} using data coming from the Planck satellite and others cosmological observations determined the $\om$ parameter region $-407.0 < \om < 175.87$ at the $95\%$ confidence level, while for positive values of this parameter they obtained $\om > 181.65$ 
at the $95\%$ confidence level. On the other hand Fabris et al. in \cite{Fabris:2005yt} using the supernovae Ia data obtained the best fit value of $\om = -1.477$. We must remember that all these limits are model dependent. In some estimations the potential of the scalar field is ignored while in others the Newtonian approximation and spherical symmetry is assumed at the starting point.

In this paper we find observational constraints on the Brans-Dicke cosmological model assuming the Robertson-Walker symmetry working at the cosmological scale. Therefore the $H^2 (a)$ relation is a starting point of our further estimations of the model parameters. The parameter 
$\om$ is hidden behind the density parameters of the Brans-Dicke modification of the Friedmann equation. The next step is to estimate the value of the density parameters from the astronomical data and compare the model with the standard cosmological model $\Lambda$CDM using information criteria. Because we treat the new model as a generalization of the $\Lambda$CDM model it is naturally to interpret a prime contribution to the $H^2 (a)$ relation as a corresponding term in the $\Lambda$CDM model.

The action for the Brans-Dicke theory \cite{Brans:1961sx} in the so-called Jordan
frame is in the following form \cite{Faraoni:book,Capozziello:book},
\begin{equation}
\label{eq:action}
S = \int\ud^{4}x\sqrt{-g}\left(\phi R
-\frac{\om}{\phi}\nabla^{\alpha}\phi\nabla_{\alpha}\phi - 
2\,V(\phi)\right) + 16\pi S_m\,,
\end{equation}
where the barotropic matter is described by
\begin{equation}
S_{m} = \int\ud^{4}x\sqrt{-g}\mathcal{L}_{m}\,,
\end{equation}
and $\om$ is a dimensionless parameter of the theory.

For the spatially flat Friedmann-Robertson-Walker metric field equations lead to the energy conservation condition
\begin{equation}
\label{eq:encon}
3H^{2} = \frac{\om}{2}\frac{\dot{\phi}^{2}}{\phi^{2}}+\frac{V(\phi)}{\phi} - 3H\frac{\dot{\phi}}{\phi} + \frac{8\pi}{\phi}\rho_{m}\,,
\end{equation}
where a dot denotes differentiation with respect to the cosmic time, and the acceleration equation
\begin{equation}
\label{eq:accel}
\begin{split}
\dot{H} = & -\frac{\om}{2}\frac{\dot{\phi}^{2}}{\phi^{2}}-\frac{1}{3+2\om}\frac{2V(\phi)-\phi V'(\phi)}{\phi} + \\ & + 2H\frac{\dot{\phi}}{\phi}-\frac{8\pi}{\phi}\rho_{m}\frac{2+\om(1+w_{m})}{3+2\om}\,,
\end{split}
\end{equation}
while the equation of motion for the scalar field is in the following form:
\begin{equation}
\ddot{\phi}+3H\dot{\phi}=2\frac{2V(\phi)-\phi V'(\phi)}{3+2\om}+8\pi\rho_{m}\frac{1-3w_{m}}{3+2\om}\,.
\end{equation}

\section{Dynamics and the Hubble function}

Using the expansion normalized variables \cite{Hrycyna:2013hla,Hrycyna:2013yia}
\begin{equation}
x\equiv\frac{\dot{\phi}}{H\phi}\,,\quad y\equiv\sqrt{\frac{V(\phi)}{3\phi}}\frac{1}{H}\,,\quad \lambda\equiv-\phi\frac{V'(\phi)}{V(\phi)}\,,
\end{equation}
the energy conservation condition \eqref{eq:encon} can be presented as
\begin{equation}
\label{eq:encon_norm}
\Omega_{m} = \frac{8\pi\rho_{m}}{3\phi H^{2}}=1+x-\frac{\om}{6}x^{2}-y^{2}\,,
\end{equation}
and the acceleration equation \eqref{eq:accel} as
\begin{equation}
\label{eq:accel_norm}
\begin{split}
\frac{\dot{H}}{H^{2}} = & 2x-\frac{\om}{2}x^{2}-\frac{3}{3+2\om}y^{2}\big(2+\lambda\big) - \\ & 
-3\left(1+x-\frac{\om}{6}x^{2}-y^{2}\right)\frac{2+\om(1+w_{m})}{3+2\om}\,.
\end{split}
\end{equation}
Then the dynamics of the Brans-Dicke theory with an arbitrary potential function and the barotropic matter content can be reduced to a three-dimensional autonomous dynamical system
\begin{subequations}
\label{eq:dynsys}
\begin{eqnarray}
\frac{\ud x}{\ud \tau} & = & -3x - x^{2} - x \frac{\dot{H}}{H^{2}} +
\frac{6}{3+2\om}y^{2}(2+\lambda) + \nonumber \\
& & +3\left(1 + x -\frac{\om}{6}x^{2} - y^{2}\right)\frac{1-3w_{m}}{3+2\om}\,,\\
\frac{\ud y}{\ud \tau} & = &  -y\left(\frac{1}{2}x(1+\lambda)+\frac{\dot{H}}{H^{2}}\right)\,, \\
\label{eq:dynsys_c}
\frac{\ud \lambda}{\ud \tau} & = &  x\lambda\Big(1-\lambda(\Gamma-1)\Big)\,,
\end{eqnarray}
\end{subequations}
where $\frac{\ud}{\ud \tau}=\frac{\ud}{\ud \ln{a}}$ and
\begin{equation}
\Gamma=\frac{V''(\phi)V(\phi)}{V'(\phi)^{2}}\,,
\end{equation}
where $()'=\frac{\ud}{\ud\phi}.$

If we assume that $\Gamma=\Gamma(\lambda)$ we are able to find critical points of the system 
\eqref{eq:dynsys} which depend on the explicit form of the $\Gamma(\lambda)$ function. In our 
previous paper \cite{Hrycyna:2013hla} we have found that for an arbitrary potential function of 
the scalar field which can be expressed by some $\Gamma(\lambda)$ function, there exist a single 
critical point ($x^{*}=0$, $y^{*}=1$, $\lambda^{*}=-2$) corresponding the de Sitter expansion.

Qualitative behavior of the solutions of the system \eqref{eq:dynsys} in the vicinity of this critical point depend on the eigenvalues of the linearization matrix calculated at this point. The eigenvalues are
\begin{equation}
\label{eq:eigval}
\begin{split}
l_{1} & = -3(1+w_{m})\,,\\
l_{2,3} & =-\frac{3}{2}\left(1\pm\sqrt{\frac{3+2\om+\delta}{3+2\om}}\right)\,,
\end{split}
\end{equation}
where $\delta$ parameter is defined as
\begin{equation}
\delta = \frac{8}{3}\lambda^{*}\big(1-\lambda^{*}(\Gamma(\lambda^{*})-1))\big)=\frac{16}{3}\big(1-2\,\Gamma^{*}\big)\,,
\end{equation}
and depends on the second derivative of the potential function at the de Sitter state. Note that for a quadratic potential function $V(\phi)\propto \phi^{2}$ we have $\Gamma=\frac{1}{2}$ which leads to $\delta=0$. Simple inspection of the eigenvalues gives that in this case one of them vanishes giving rise to degenerated critical point and structurally unstable system \cite{Hrycyna:2013yia,Szydlowski:2013sma}.

For $w_{m}>-1$ the critical point corresponding to the de Sitter expansion is stable when $\frac{\delta}{3+2\om}<0$ and represents a saddle-type critical point otherwise. The stable case can be further divided in to two cases corresponding to a stable node for $-1<\frac{\delta}{3+2\om}<0$ and a stable focus for $\frac{\delta}{3+2\om}<-1$. In the most general case, from \eqref{eq:eigval}, one can distinguish two cases, the first one when the eigenvalues of the linearization matrix are purely real and the second when the eigenvalues have a nonzero imaginary part. The case with purely imaginary eigenvalues is excluded in our case.

From now on we assume that we include only the baryonic matter, $\Omega_{m}=\Omega_{bm}$ with the equation of state parameter $w_{m}=0$.

In Appendix~\ref{app_a} we presented the linearized solutions in the vicinity of the de Sitter state for two types of behavior.

In the first case, characterized by the purely real eigenvalues, we make the following substitution
\begin{equation}
\label{eq:om_1}
\frac{\delta}{3+2\om}=\frac{4}{9}n(n-3)\,,
\end{equation}
and the eigenvalues of the linearization matrix at the de Sitter state are
\begin{equation}
l_{1}=-3\,, \quad l_{2}=-n\,, \quad l_{3}=-3+n\,,
\end{equation}
where for $0<n<\frac{3}{2}$ we have a stable node critical point and for $n<0$ a saddle type. Note that the case of $n=0$ or $n=3$, which corresponds to $\delta=0$ or $\om=\infty$, is excluded from our investigations as this case leads to degenerated critical point and structurally unstable system  \cite{Hrycyna:2013yia,Szydlowski:2013sma}.

Now, using the linearized solutions \eqref{eq:sol_lin} and the acceleration equation \eqref{eq:accel_norm} up to linear terms in initial conditions we obtain the following Hubble function,
\begin{widetext}
\begin{equation}
\label{eq:1}
\left(\frac{H(a)}{H(a_{0})}\right)^{2} = \Omega_{\Lambda,0} +
\Omega_{M,0}\left(\frac{a}{a_{0}}\right)^{-3} +
\Omega_{n,0}\left(\frac{a}{a_{0}}\right)^{-n} +
\Omega_{3n,0}\left(\frac{a}{a_{0}}\right)^{-3+n} \,,
\end{equation}
where
\begin{subequations}
\begin{align}
\label{eq:M_lin}
\Omega_{M,0}  = &\bigg(1- 
\frac{16}{3\delta}\bigg)\,\Omega_{bm,0}\,,\\
\Omega_{n,0} =  &
\frac{n+1}{3\delta(2n-3)}\bigg(4n\,\Omega_{bm,i} -3\delta\,
\Delta x+8(n-3)\Delta\lambda\bigg)\left(\frac{a_{0}}{a^{(i)}}\right)^{-n}\,,\\
\Omega_{3n,0} = & 
\frac{n-4}{3\delta(2n-3)}\bigg(-4(n-3)\,\Omega_{bm,i}
-3\delta\,\Delta x -8n\Delta\lambda\bigg) \left(\frac{a_{0}}{a^{(i)}}\right)^{-3+n} \,,
\end{align}
\label{eq:Omega_1}
\end{subequations}
and
\begin{equation} 
\Omega_{\Lambda,0} = 1- \Omega_{M,0}-\Omega_{n,0}-\Omega_{3n,0}\,,
\end{equation}
\end{widetext}
where $\Delta x = x^{(i)}$, $\Delta y = y^{(i)}-1$, and $\Delta\lambda= \lambda^{(i)}+2$ are the initial conditions in the vicinity of the de Sitter state and $a^{(i)}$, $a_{0}$ are the initial and the present value of the scale factor. Up to linear terms in the initial conditions, from \eqref{eq:encon_norm} we have $\Omega_{bm,i}=\Delta x - 2\Delta y$.

In our further investigation the model described by the Hubble function \eqref{eq:1} together with $0<n<\frac{3}{2}$, we denote as ``model 1a'' (the de Sitter state is the critical point of a stable node type), while the model with $n<0$ we denote as ``model 1b'' (the de Sitter state is a saddle-type critical point).

For the second type of behavior in the vicinity of the de Sitter state we make the following substitution,
\begin{equation}
\label{eq:om_2}
\frac{\delta}{3+2\om} = -\frac{1}{9}(9+4n^{2})\,,
\end{equation}
and the eigenvalues of the linearization matrix at the de Sitter state are
\begin{equation}
l_{1}=-3\,,\quad l_{2}=-\frac{3}{2}-\mathbbmtt{i} n\,,\quad l_{3}=-\frac{3}{2}+\mathbbmtt{i} n\,.
\end{equation}
From the solutions \eqref{eq:sol_osc} and the acceleration equation \eqref{eq:accel_norm}, and again, up to linear terms in initial conditions we obtain the following Hubble function, 
\begin{widetext}
\begin{equation}
\label{eq:2}
\left(\frac{H(a)}{H(a_{0})}\right)^{2}  =  \Omega_{\Lambda,0} +  
\Omega_{M,0}\left(\frac{a}{a_{0}}\right)^{-3} 
+\left(\frac{a}{a_{0}}\right)^{-3/2}
\Bigg(
\Omega_{cos,0}\,\cos{\left(n\,\ln{\bigg(\frac{a}{a_{0}}\bigg)}\right)} +
\Omega_{sin,0}\,\sin{\left(n\,\ln{\bigg(\frac{a}{a_{0}}\bigg)}\right)}
\Bigg)\,,
\end{equation}
where
\begin{subequations}
\begin{align}
\label{eq:M_osc}
\Omega_{M,0} = & \bigg(1-\frac{16}{3\delta}\bigg)\Omega_{bm,0}\,,\\
\Omega_{cos,0} = & 
\frac{1}{3\delta}\bigg(16\,\Omega_{bm,i} 
-3\delta\,\Delta x + 8 \Delta\lambda\bigg)
\left(\frac{a_{0}}{a^{(i)}}\right)^{-3/2}\cos{\left(n\,\ln{\bigg(\frac{a_{0}}{a^{(i)}}\bigg)}
\right)} + \nonumber \\
 + &\frac{1}{6\delta n}\bigg(2(4n^{2}-15)
\,\Omega_{bm,i} 
+ 15\delta\,\Delta x + 4(4n^{2}+15) \Delta\lambda
\bigg)\left(\frac{a_{0}}{a^{(i)}}\right)^{-3/2}\sin{\left(n\,\ln{\bigg(\frac{a_{0}}{a^{(i)}}
\bigg)}\right)}\,,\\
\Omega_{sin,0} = &\frac{1}{6\delta n}\bigg(2(4n^{2}-15)\,\Omega_{bm,i} 
+ 15\delta\,\Delta x + 4(4n^{2}+15) \Delta\lambda
\bigg)\left(\frac{a_{0}}{a^{(i)}}\right)^{-3/2}\cos{\left(n\,\ln{\bigg(\frac{a_{0}}{a^{(i)}}
\bigg)}\right)} - \nonumber \\
-&\frac{1}{3\delta}\bigg(16\,\Omega_{bm,i} 
-3\delta\,\Delta x + 8 \Delta\lambda\bigg)
\left(\frac{a_{0}}{a^{(i)}}\right)^{-3/2}\sin{\left(n\,\ln{\bigg(\frac{a_{0}}{a^{(i)}}\bigg)}
\right)}\,. 
\end{align}
\end{subequations}
and
\begin{equation}
\Omega_{\Lambda,0} = 1-\Omega_{M,0}-\Omega_{cos,0}\,,
\end{equation}
\end{widetext}
where $\Delta x = x^{(i)}$, $\Delta y = y^{(i)}-1$, and $\Delta\lambda= \lambda^{(i)}+2$ are the initial conditions in the vicinity of the de Sitter state and $a^{(i)}$, $a_{0}$ are the initial and the present value of the scale factor. Up to linear terms in the initial conditions, from \eqref{eq:encon_norm} we have $\Omega_{bm,i}=\Delta x - 2\Delta y$.

The model described by the Hubble function \eqref{eq:2} we denote as ``model 2'' (the de Sitter state corresponds to the critical point of a stable focus type).

Note that for the Hubble functions \eqref{eq:1} and \eqref{eq:2}, when the parameter $\delta$ in \eqref{eq:M_lin} and \eqref{eq:M_osc} is negative $\delta<0$, the density parameter of the matter content $\Omega_{M,0}$ is larger than the density parameter of the matter included in the model by hand $\Omega_{bm,0}$.

Additionally the $\Lambda$CDM model is nested within both Hubble functions \eqref{eq:1} and \eqref{eq:2}, i.e. carefully choosing the initial conditions for the linearized solutions 
\begin{equation}
\Delta x =\frac{4}{\delta}\Omega_{bm,i}\,, \qquad \Delta\lambda = -\frac{1}{2}\Omega_{bm,i}\,,
\end{equation}
where up to linear terms in initial conditions $\Omega_{bm,i}=\Delta x -2\Delta y$, then in \eqref{eq:1} we have $\Omega_{n,0}=\Omega_{3n,0}=0$ and in \eqref{eq:2} we have 
$\Omega_{cos,0}=\Omega_{sin,0}=0$ and the resulting form of the Hubble function is
\begin{equation}
\left(\frac{H(a)}{H(a_{0})}\right)^{2} \approx 1-\Omega_{M,0} + \Omega_{M,0}\left(\frac{a}{a_{0}}\right)^{-3}\,,
\end{equation}
where
\begin{equation}
\Omega_{M,0} = \bigg(1-\frac{16}{3\delta}\bigg)\Omega_{bm,0}\,.
\end{equation}
This Hubble function describes the $\Lambda$CDM model with direct interpretation of the 
second term in the brackets as proportional to density parameter of the dark matter in the model
\begin{equation}
\Omega_{dm,0} = -\frac{16}{3\delta}\Omega_{bm,0}\,.
\end{equation} 

\section{Observational constraints}

To estimate the parameters of the models we used modified for our purposes, publicly available \textsc{CosmoMC} source code \cite{cosmomc,Lewis:2002ah} with implemented nested sampling algorithm \textsc{multinest} \cite{Feroz:2007kg,Feroz:2008xx,Feroz:2013hea}. We kept fixed present values of the Hubble function $H_{0}= 67.4\,\, \text{Mpc/km/s}$ and the baryonic matter density parameter $\Omega_{bm,0}\text{h}^{2} = 0.02207$ taken for the recent observations of the Planck satellite \cite{Ade:2013zuv}. In all investigated models we assumed a flat prior for estimated parameters in the following intervals: $\Omega_{M,0}\in(0.1;0.5)$, $\Omega_{n,0}\in(-1;1)$ , $\Omega_{3n,0}\in(-1;1)$ and the parameter $n$ for the model 1a $n\in(0;\frac{3}{2})$ and for the model 1b $n\in(-3;0)$. For the model 2 we assumed $\Omega_{sin,0}\in(-1;1)$, $\Omega_{cos,0}\in(-1;1)$ and $n\in(0;5)$.

We used observational data of 580 supernovae type Ia the so called \textsc{Union2.1} compilation \cite{Suzuki:2011hu}, 31 observational data points of Hubble function from \cite{Jimenez:2001gg,Simon:2004tf,Gaztanaga:2008xz,Stern:2009ep,Moresco:2012jh,Busca:2012bu,Zhang:2012mp,Blake:2012pj,Chuang:2012qt,Anderson:2013oza} collected in \cite{Chen:2013vea}, the measurements of BAO (baryon acoustic oscillations) from Sloan Digital Sky Survey (SDSS-III) combined with 2dF Galaxy Redshift Survey (2dFGRS) \cite{Eisenstein:2005su,Percival:2009xn,Eisenstein:2011sa,Ahn:2013gms}, The 6dF Galaxy Survey (6dFGS) \cite{Jones:2009yz,Beutler:2011hx},  WiggleZ Dark Energy Survey \cite{Drinkwater:2009sd,Blake:2011en,Blake:2011wn} and information coming from determinations of Hubble function using Alcock-Paczy\'{n}ski test \cite{Alcock:1979mp,Blake:2011ep}.

In this paper the starting point was the Hubble functions obtained from the linearized solutions in the vicinity of the de Sitter state. Such solutions have a limited range of applicability and cannot be prolonged up to arbitrary values of the scale factor, so we did not apply obtained Hubble functions to observational data coming from large redshifts.
The observational data coming from the CMB are beyond the scope of applicability of the obtained Hubble functions.

The likelihood function for the supernovae data is defined by
\begin{equation}
L_{SN} \propto \exp \left[ - \sum_{i,j}(\mu_{i}^{\mathrm{obs}} - \mu_{i}^{\mathrm{th}}) \mathbb{C}_{ij}^{-1} (\mu_{j}^{\mathrm{obs}} - \mu_{j}^{\mathrm{th}})\right] \, ,\label{sn_likelihood}
\end{equation}
where $\mathbb{C}_{ij}$ is the covariance matrix with the systematic errors, $\mu_{i}^{\mathrm{obs}}=m_{i}-M$ is the distance modulus, 
$\mu_{i}^{\mathrm{th}}=5\log_{10}D_{Li} + \mathcal{M}=5\log_{10}d_{Li} + 25$, $\mathcal{M}=-5\log_{10}H_{0}+25$ and $D_{Li}=H_{0}d_{Li}$, where $d_{Li}$ is the luminosity distance which is given by $d_{Li}=(1+z_{i})c\int_{0}^{z_{i}} \frac{dz'}{H(z')}$ (with the assumption $k=0$). 

For $H(z)$ the likelihood function is given by
\begin{equation}
L_{H(z)} \propto \exp \left[ - \sum_i\frac{\left(H^{\mathrm{th}}(z_i)-H^{\mathrm{obs}}_i\right)^2}{2 \sigma_i^2} \right ],
\label{hz_likelihood}
\end{equation}
where $H^{\mathrm{th}}(z_i)$ denotes the theoretically estimated Hubble function, $H^{\mathrm{obs}}_i$ is observational data.

For BAO A parameter the likelihood function is defined as
\begin{equation}
L_{BAOA} \propto \exp \left[ - \sum_{i,j}(A^{\mathrm{th}}(z_i)-A^{\mathrm{obs}}_i) \mathbb{C}_{ij}^{-1} (A^{\mathrm{th}}(z_j)-A^{\mathrm{obs}}_j)\right] \, ,\label{baoa_likelihood}
\end{equation}
where $\mathbb{C}_{ij}$ is the covariance matrix with the systematic errors, $A^{\mathrm{th}}(z_i)=\sqrt{\Omega_{m,0}} \left (\frac{H(z_i)}{H_{0}} \right ) ^{-\frac{1}{3}} \left [ \frac{1}{z_i} \int_{0}^{z_i}\frac {H_0}{H(z)} dz\right]^{\frac{2}{3}}$.

The likelihood function for the rest of BAO data is characterized by
\begin{equation}
L_{BAO} \propto \exp \left[ - \sum_{i,j}\left(d^{\mathrm{th}}(z_i)-d^{\mathrm{obs}}_i\right) \mathbb{C}_{ij}^{-1} \left(d^{\mathrm{th}}(z_j)-d^{\mathrm{obs}}_j\right)\right] \, ,\label{bao_likelihood}
\end{equation}
where $\mathbb{C}_{ij}$ is the covariance matrix with the systematic errors, $d^{\mathrm{th}}(z_i)\equiv r_s(z_d) \left[(1+z_i)^2 D_A^2(z_i)\frac{cz_i}{H(z_i)} \right ]^{-\frac{1}{3}}$, $r_s(z_d)$ is the sound horizon at the drag epoch and $D_A$ is the angular diameter distance.

And finally, the likelihood function for the information coming from Alcock-Paczy\'{n}ski test is given by
\begin{equation}
L_{AP} \propto \exp \left[ - \sum_i\frac{\left(AP^{\mathrm{th}}(z_i)-AP^{\mathrm{obs}}_i\right)^2}{2 \sigma_i^2} \right ],
\label{ap_likelihood}
\end{equation}
where: $AP^{\mathrm{th}}(z_i)\equiv \frac{H(z_i)}{H_0 (1+z_i)}$.

The total likelihood function $L_{TOT}$ is defined as
\begin{equation}
L_{TOT}=L_{SN}L_{H(z)}L_{BAOA}L_{BAO}L_{AP}.
\label{total_likelihood}
\end{equation}

The mean of marginalized posterior PDF with $68 \%$ confidence level and the values of the joined posterior probabilities of the parameters for all investigated models are gathered in Table~\ref{tab_values}. 

\begin{center}
\begin{figure}
\includegraphics[scale=0.45]{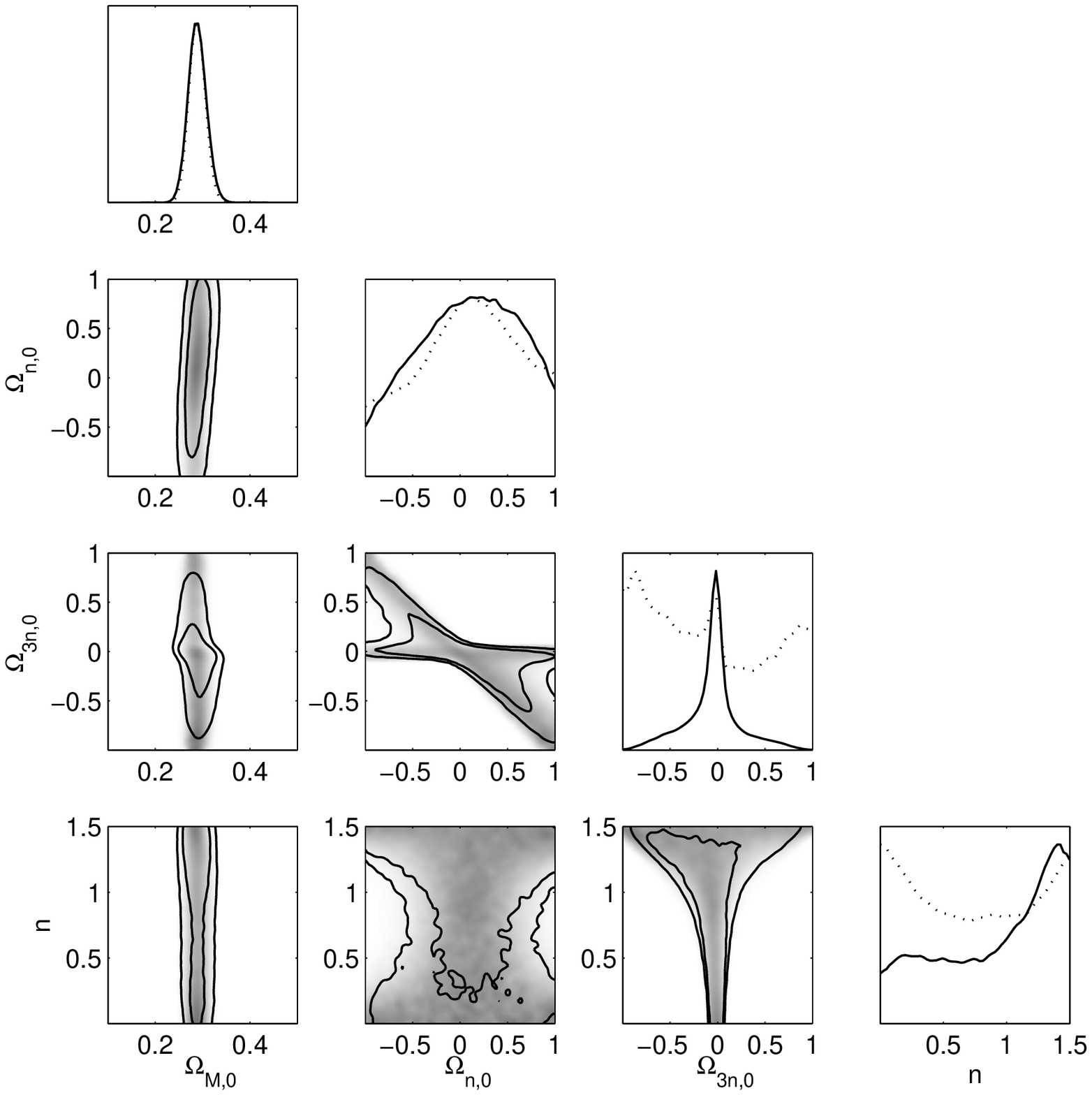}
\caption{Posterior constraints for investigated model 1a (linear approach to the de Sitter state). One-dimensional plots: solid lines denote fully marginalized probabilities, dotted lines show mean likelihood. Two-dimensional plots: solid lines denote 68 \% and 95 \% credible intervals of fully marginalized probabilities, the colors illustrate mean likelihood of the sample. Estimations were made using   \emph{Union2.1+H(z)+Alcock-Paczy\'{n}ski+BAO} data set. For the numerical results see Table~\ref{tab_values}.}
\label{fig:1}
\end{figure}
\end{center}

\begin{center}
\begin{figure}
\includegraphics[scale=0.45]{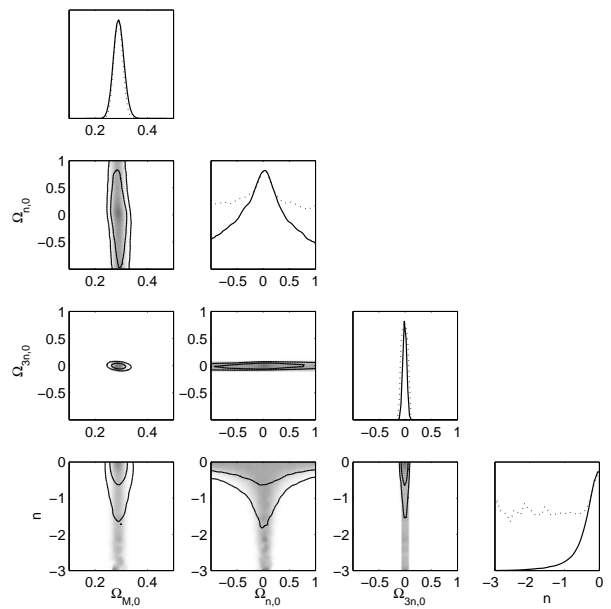}
\caption{Posterior constraints for investigated model 1b (transient de Sitter evolution). One-dimensional plots: solid lines denote fully marginalized probabilities, dotted lines show mean likelihood. Two-dimensional plots: solid lines denote 68 \% and 95 \% credible intervals of fully marginalized probabilities, the colors illustrate mean likelihood of the sample. Estimations were made using   \emph{Union2.1+H(z)+Alcock-Paczy\'{n}ski+BAO} data set. For the numerical results see Table~\ref{tab_values}.}
\label{fig:2}
\end{figure}
\end{center}

\begin{center}
\begin{figure}
\includegraphics[scale=0.45]{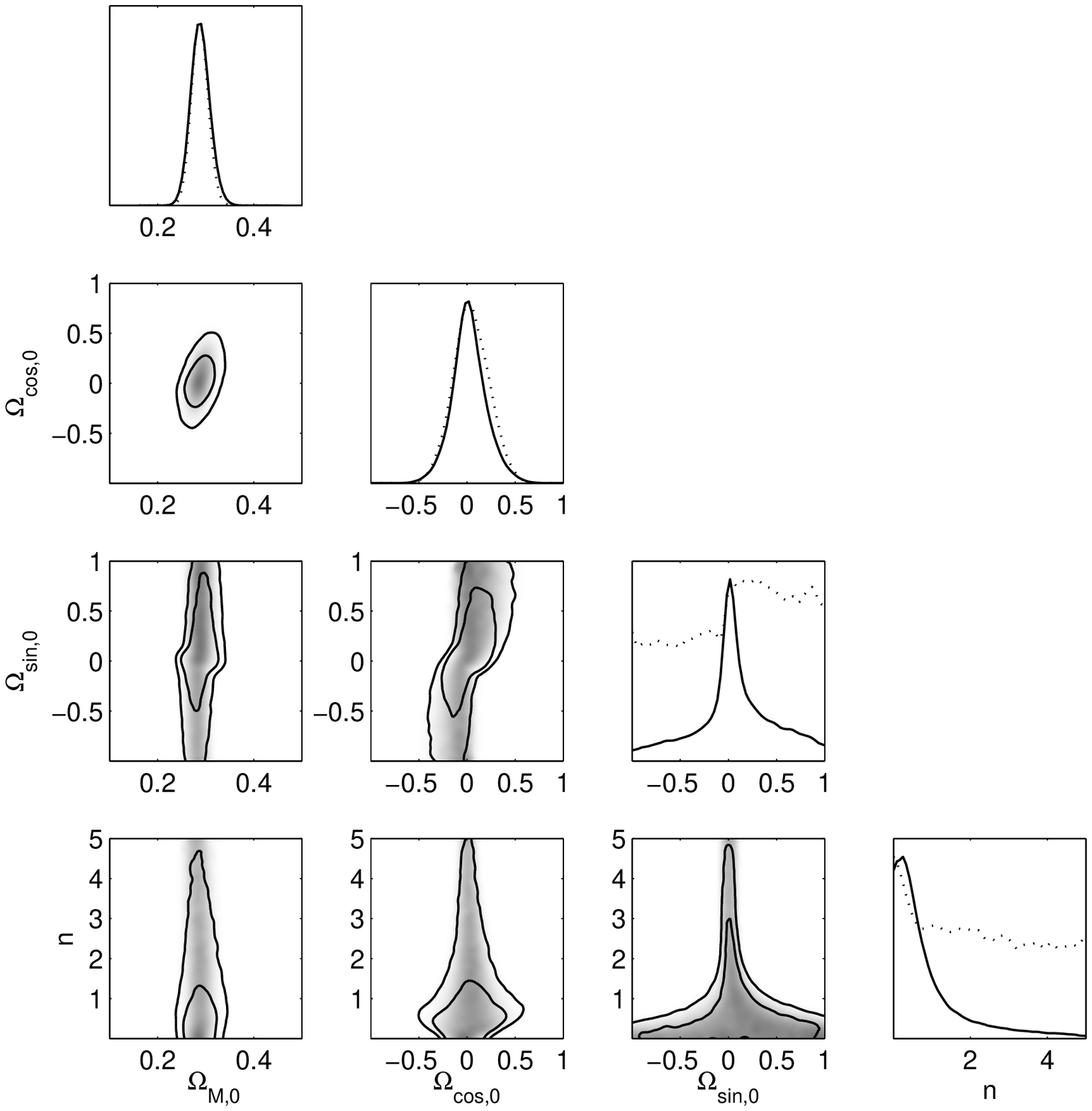}
\caption{Posterior constraints for investigated model 2 (oscillatory approach to the de Sitter state). One-dimensional plots: solid lines denote fully marginalized probabilities, dotted lines show mean likelihood. Two-dimensional plots: solid lines denote 68 \% and 95 \% credible intervals of fully marginalized probabilities, the colors illustrate mean likelihood of the sample. Estimations were made using  \emph{Union2.1+H(z)+Alcock-Paczy\'{n}ski+BAO} data set. For the numerical results see Table~\ref{tab_values}.}
\label{fig:3}
\end{figure}
\end{center}

The posterior constraints for investigated models are given in Figs.~\ref{fig:1}, \ref{fig:2}, and \ref{fig:3}. On the one-dimensional plots the solid lines denote fully marginalized probabilities and the dotted lines show mean likelihood. On the two-dimensional plots the solid lines denote 68\% and 95\% credible intervals of fully marginalized probabilities while the colors illustrate mean likelihood of the sample used.

\begin{table}
\centering
\caption{Mean of marginalized posterior PDF with $68 \%$ confidence level for the parameters of the models. In the brackets are shown parameter's values of joined posterior probabilities. Estimations were made using \emph{Union2.1}, \emph{H(z)}, \emph{Alcock-Paczy\'{n}ski} and \emph{BAO} data sets.}
\label{tab_values}
\begin{ruledtabular}
\renewcommand{\arraystretch}{2}
\begin{tabular}{|c|c|c|}
  & \emph{Union2.1+H(z)+AP} & \emph{Union2.1+H(z)+AP+BAO}\\
  \hline
  \multicolumn{3}{|c|}{model 1a}\\
  \hline
    $\Omega_{M,0}$ 
    & $0.2788^{+0.0939}_{-0.0931} (0.1328)$ & $0.2887^{+0.0180}_{-0.0178} (0.2882)$ \\ 
    $\Omega_{n,0}$ 
    & $0.0083^{+0.6486}_{-0.6464} (-0.6933)$ & $0.0760^{+0.5673}_{-0.5795} (0.9427)$ \\ 
    $\Omega_{3n,0}$ 
    & $0.0144^{+0.2929}_{-0.287} (0.2562)$ & $-0.0517^{+0.1988}_{-0.2468} (-0.0149)$ \\
    $n $
    & $0.7305^{+0.5324}_{-0.5185} (0.4150)$ & $0.9073^{+0.4704}_{-0.5720} (0.0210)$ \\
  \hline
  \multicolumn{3}{|c|}{model 1b}\\
  \hline
    $\Omega_{M,0}$ 
    & $0.3053^{+0.1296}_{-0.1320} (0.3056)$ & $0.2903^{+0.0190}_{-0.0191} (0.2873)$ \\
    $\Omega_{n,0}$ 
    & $0.0090^{+0.6093}_{-0.6137} (0.0720)$ & $-0.0121^{+0.5182}_{-0.5264} (0.0179)$ \\
    $\Omega_{3n,0}$ 
    & $-0.0184^{+0.0938}_{-0.0922} (-0.0043) $ & $-0.0088^{+0.0170}_{-0.0181} (-0.0025)$ \\
    $n$ 
    & $-0.3585^{+0.2881}_{-0.2832} (-1.7160)$ & $-0.4012^{+0.3229}_{-0.3082} (-1.3276)$ \\
  \hline
  \multicolumn{3}{|c|}{model 2}\\
  \hline
    $\Omega_{M,0}$ 
    & $0.2888^{+0.0187}_{-0.0187} (0.2891)$ & $0.2723^{+0.0642}_{-0.0674} (0.2200)$ \\ 
    $\Omega_{cos,0}$ 
    & $0.0229^{+0.1676}_{-0.1626} (0.0540)$ & $0.0034^{+0.2497}_{-0.2530} (-0.1773)$ \\
    $\Omega_{sin,0}$ 
    & $0.0688^{+0.4189}_{-0.3662} (0.5060)$ & $-0.0318^{+0.5839}_{-0.5709} (-0.7339)$ \\
    $n$ 
    & $0.9295^{+0.7859}_{-0.7538} (0.1981)$ & $0.9153^{+0.6141}_{-0.6722} (0.5139)$ \\
  \hline
  \multicolumn{3}{|c|}{$\Lambda$CDM model}\\
  \hline
    $\Omega_{M,0}$ 
    & $0.2792^{+0.0228}_{-0.0229} (0.2772)$ & $0.2850^{+0.0164}_{-0.0164} (0.2839)$ \\
\end{tabular}
\end{ruledtabular}
\end{table}

In order to discriminate between models we used twice of the natural logarithm of the Bayes factor of two models defined as 
\begin{equation}
2\ln{B_{0i}} = 2\ln{\frac{E_{0}}{E_{i}}}\,,
\end{equation}
which is proportional to the ratio of the evidence of the base model $E_{0}$ and the evidence of the model investigated $E_{i}$.

This quantity can be interpreted as a evidence in favor of the base model with subscript 
``$0$''. For $2>2\ln{B_{0i}}>0$ the evidence is not worth a bare mention, for $6>2\ln{B_{0i}}>2$  is positive, for $10>2\ln{B_{0i}}>6$ is strong and when $2\ln{B_{0i}}>10$ the evidence is very strong in favor of model ``$0$'' (or very strong evidence against model ``$i$'') \cite{Kass:1995}.

The values of twice the natural logarithm of the Bayes factor of 
models 1a, 1b, and 2 with respect to the $\Lambda$CDM model 
are gathered in Table II. Using \emph{Union2.1+H(z)+Alcock-Paczy\'{n}ski} data set we obtain a positive evidence in favor of the $\Lambda$CDM model over the models under considerations. The \emph{Union2.1+H(z)+Alcock-Paczy\'{n}ski+BAO} data set 
gives positive evidence of the $\Lambda$CDM model over the model 1a while the models 1b and 2 are strongly disfavored (or, equivalently, the strong evidence in favor of the $\Lambda$CDM as
compared to the two models considered). 

Calculating twice the natural logarithm of the Bayes factor between the models under investigations we obtain, for models 1a and 1b : $2\ln{B_{1a1b}}=3.74\pm0.35$, for models 1a and 2 : $2\ln{B_{1a2}}=1.50\pm0.32$ and for models 2 and 1b : $2\ln{B_{21b}}=2.24\pm0.38$. These values give a positive evidence in favor of the models 1a and 2 with respect to the model 1b. The models with the de Sitter state in form of an attractor of the system are favored in light of the used observational data.

The Bayesian statistical analysis crucially depends on the choice of the parameters priors. The models under considerations were obtained from the linearized solutions to dynamics in the vicinity of the de Sitter state and hence the $\Omega_{i,0}$ in the Hubble functions can not be arbitrary large as they depend linearly on the initial conditions. For larger regions on the parameter priors the evidence of the model will be worse but we must remember that in these models the main contribution to the Hubble functions are terms similar to the terms in the $\Lambda$CDM model. Restricting the allowed range for parameters one can obtain models which are virtually indistinguishable from the standard model.

\begin{table}
\centering
\caption{Values of the evidence and the Bayes factor
(with respect to $\Lambda$CDM model) for \emph{Union2.1+H(z)+Alcock-Paczy\'{n}ski} and \emph{Union2.1+H(z)+Alcock-Paczy\'{n}ski+BAO} data sets.}
\label{aic_bic}
\renewcommand{\arraystretch}{2}
\begin{tabular}{|c|c|c|c|}
  \hline\hline
  \multicolumn{3}{|c|}{\emph{Union2+H(z)+AP}}\\
  \hline
    & evidence $\ln{E_{i}}$ 
    &  $2\ln{B_{0i}}$\\
  \hline
model 1a    & $-285.55\pm0.10$ 
& $2.73\pm0.25$ 
\\
model 1b    & $-286.98\pm0.12$ 
& $5.60\pm0.28$ 
\\
model 2     & $-286.61\pm0.10$ 
& $4.85\pm0.26$ 
\\
$\Lambda$CDM & $-284.18\pm0.08$ 
& $0$ 
\\
  \hline\hline
  \multicolumn{3}{|c|}{\emph{Union2+H(z)+AP+BAO}}\\
  \hline
      &evidence $\ln{E_{i}}$ 
      &
      $2\ln{B_{0i}}$\\
  \hline
model 1a    & $-288.02\pm0.11$ 
& $5.16\pm0.27$ 
\\
model 1b    & $-289.89\pm0.14$ 
& $8.89\pm0.33$ 
\\
model 2     & $-288.77\pm0.12$ 
& $6.65\pm0.30$ 
\\
$\Lambda$CDM  & $-285.44\pm0.09$ 
& $0$ 
\\
 \hline\hline
\end{tabular}
\end{table}

\section{Derived quantities}

In the previous section we have estimated values of the unknown parameters of the models. The two pairs of parameters ($\Omega_{n,0}$,$\Omega_{3n,0}$) and ($\Omega_{sin,0}$,$\Omega_{cos,0}$) depend on the initial conditions of the phase space variables. Knowing the values of the parameters of the models one is able to obtain complete information about the present state of the phase space variables. 

The action integral \eqref{eq:action} gives the effective gravitational coupling in Brans-Dicke theory as an inverse of the scalar field
\begin{equation}
\label{eq:34}
G_{\text{eff}} = \frac{1}{\phi}\,.
\end{equation}
However, for the spherically symmetric solutions in the Brans-Dicke theory in Cavendish-like experiments we have \cite{EspositoFarese:2000ij, Damour:1992we, Damour:1995kt}
\begin{equation}
\label{eq:35}
G_{\text{eff}}=\frac{1}{\phi}\frac{4+2\om}{3+2\om}\,.
\end{equation}
We have to remember that this quantity is defined in the context of the parametrized post-Newtonian (PPN) formalism \cite{Will:book} only suitable for the solar system tests where a spherical symmetry of solutions is assumed, and not for our cosmological considerations of the background evolution only \cite{Faraoni:book}.

The variation of the effective gravitational coupling in the Brans-Dicke theory can be directly connected with the cosmological evolution of the scalar field
\begin{equation}
\frac{\dot{G}_{\text{eff}}}{G_{\text{eff}}} = -\frac{\dot{\phi}}{\phi}\,,
\end{equation}
which can be obtained either from \eqref{eq:34} or \eqref{eq:35}. Thus we have a direct interpretation of the present value of the phase space variable $x$
\begin{equation}
x(a_{0}) = \frac{\dot{\phi}}{H\phi}\bigg|_{0} = -\frac{\dot{G}}{H\,G}\bigg|_{0}\,,
\end{equation}
where in $G$ we omitted the subscript for simplicity. For the remaining two phase space variables we have
\begin{equation}
y(a_{0}) = \sqrt{\frac{V(\phi_{0})}{3\phi_{0}}}\frac{1}{H_{0}}\,, \qquad 
\lambda(a_{0}) = -\phi_{0}\frac{V'(\phi_{0})}{V(\phi_{0})}\,,
\end{equation}
where the first quantity is proportional to the value of the scalar field potential function at the present epoch and the second one gives the slope of the potential function at the present epoch. 

The dynamical system analysis enables us to find the asymptotic values of the phase space variables while the linearized solutions give us an opportunity to find its present time values. 
From Eqs.~\ref{eq:sol_lin} for the models 1a and 1b we can derive
\begin{subequations}
\begin{align}
\label{eq:x_lin}
\begin{split}
x(a_{0}) = \,& \frac{3}{4}\big(\Omega_{bm,0}-\Omega_{M,0}\big) - \\& -\frac{n}{n+1}\Omega_{n,0} - \frac{n-3}{n-4}\Omega_{3n,0}\,,
\end{split}\\
\begin{split}
y(a_{0}) = \,& 1- \frac{1}{8}\big(\Omega_{bm,0}+3\Omega_{M,0}\big) -\\& - \frac{1}{2}\frac{n}{n+1}\Omega_{n,0} - \frac{1}{2}\frac{n-3}{n-4}\Omega_{3n,0} \,,
\end{split}\\
\label{eq:l_lin}
\begin{split}
\lambda(a_{0}) = \,& -2 -\frac{1}{2}\Omega_{bm,0} -\\ & - \frac{3}{8}\frac{1}{n+1}\Omega_{n,0} - \frac{3}{8}\frac{1}{n-4}\Omega_{3n,0} \,,
\end{split}
\end{align}
\end{subequations}
while from \ref{eq:sol_osc} for the model 2 we find
\begin{subequations}
\begin{align}
\label{eq:x_osc}
\begin{split}
x(a_{0}) = \,& \frac{3}{4}(\Omega_{bm,0}-\Omega_{M,0})+ \\ &+\frac{2}{4n^{2}+25}\big(5\, \Omega_{cos,0}+2 n \,\Omega_{sin,0}\big) -\Omega_{cos,0}\,,
\end{split}\\
\begin{split}
y(a_{0}) = \,& 1-\frac{1}{8}\big(\Omega_{bm,0}+3\Omega_{M,0}\big) + \\ & + \frac{1}{4n^{2}+25}\big(5\, \Omega_{cos,0}+2 n \,\Omega_{sin,0}\big) -\frac{1}{2}\Omega_{cos,0}\,,
\end{split}\\
\label{eq:l_osc}
\begin{split}
\lambda(a_{0}) = \,& -2-\frac{1}{2}\Omega_{bm,0}+ \\ & + \frac{4}{4n^{2}+25}\frac{\Omega_{bm,0}}{\Omega_{bm,0}-\Omega_{M,0}}\big(5\, \Omega_{cos,0}+2 n \,\Omega_{sin,0}\big)\,.
\end{split}
\end{align}
\end{subequations}
These equations express the interrelation between the parameters estimated in the Hubble function and the phase space state of the dynamical system under considerations. 

In Table~\ref{tab:3} we have gathered the present values of the phase space variables $x(a_{0})$, $y(a_{0})$ and $\lambda(a_{0})$ calculated for the mean of marginalized posterior PDF with 68\% confidence level for the parameters of the models. 
The errors for a given quantity where calculated as minimal and maximal value of the quantity within $1\sigma$ intervals of the estimated parameters.

\begin{center}
\begin{figure}
\includegraphics[scale=0.3]{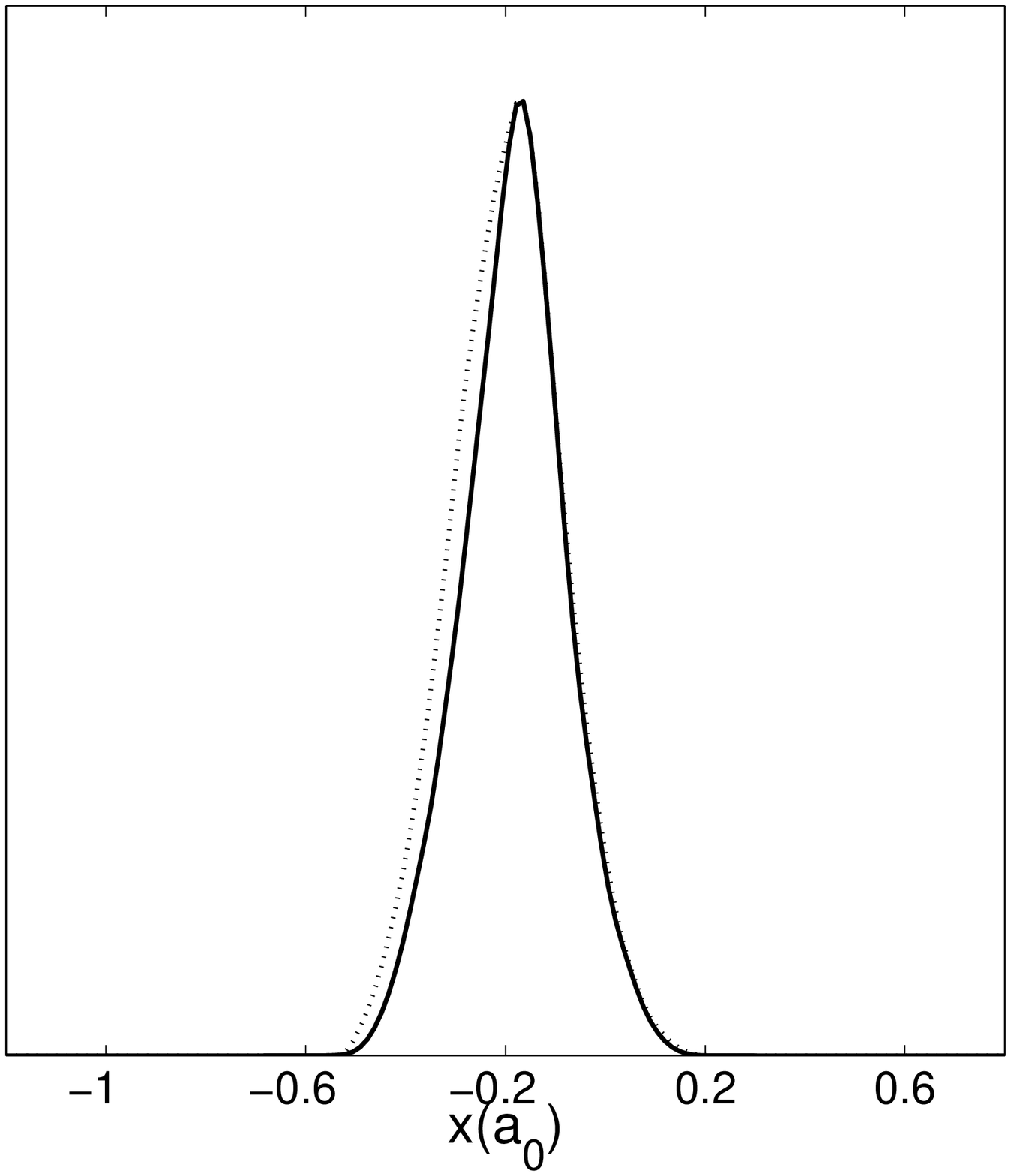}\\
\includegraphics[scale=0.3]{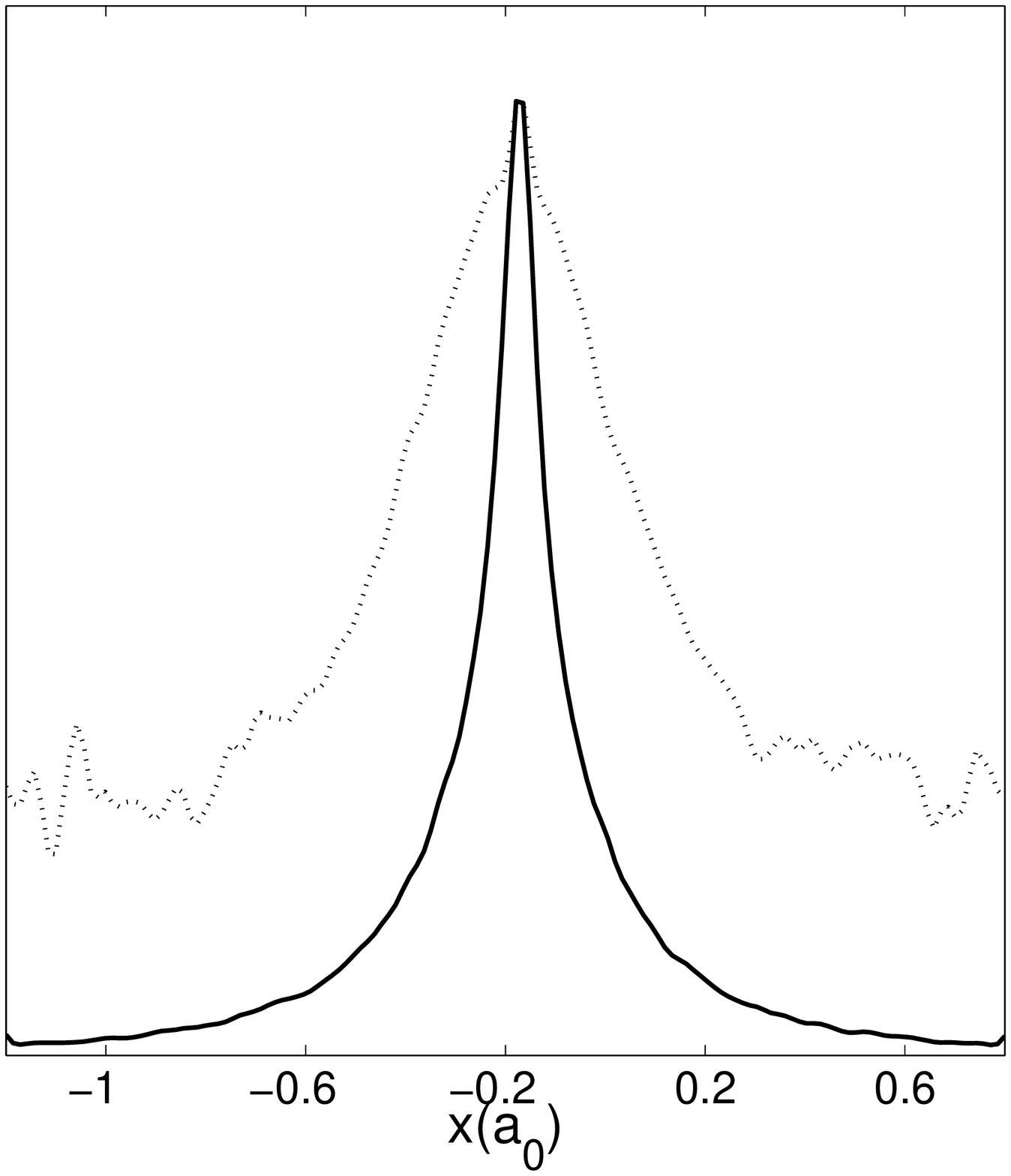}\\
\includegraphics[scale=0.3]{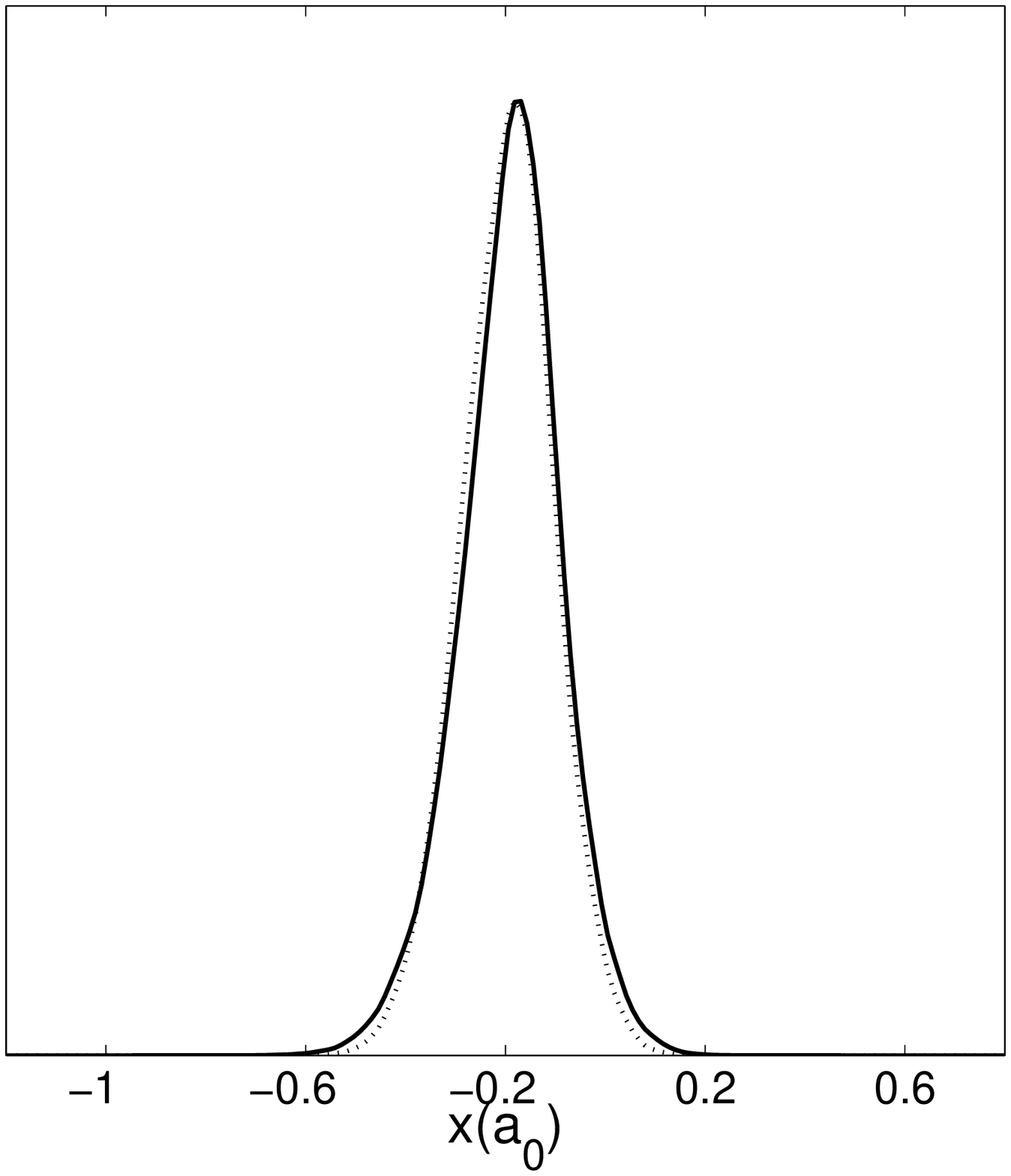}
\caption{The fully marginalized probabilities (solid lines) and mean likelihood (dotted lines) for the present value of the phase space variable $x(a_{0})$ calculated for the models $1a$ (top) and $1b$ (middle) from Eq.~\eqref{eq:x_lin} and for the model $2$ (bottom) from Eq.~\eqref{eq:x_osc}. The maximal probability is located at the negative values indicating that at present the effective gravitational coupling constant increases.}
\label{fig:4}
\end{figure}
\end{center}

In Fig.~\ref{fig:4} we present the fully marginalized probabilities (solid lines) and mean likelihood (dotted lines) for the present value of the phase space variable $x(a_{0})$ which is directly connected with time variation of the effective gravitational coupling constant. We have that at present epoch
\begin{equation}
\frac{\dot{G}}{G}\bigg|_{0} = - x(a_{0}) H_{0}\,,
\end{equation}
which indicates that for every investigated model this quantity is positive, and thus, the value of the effective gravitational coupling constant increases during the evolution of universe. This observation can indicate the weakening the strength of gravity at early times and might be the reason for the low entropy of the early universe \cite{Greene:2009tt}.

\begin{table}
\centering
\caption{The present values of the phase space variables $x(a)$, $y(a)$, and $\lambda(a)$ calculated for the mean of marginalized posterior PDF with 68\% confidence level for the parameters of the models.}
\label{tab:3}
\begin{ruledtabular}
\renewcommand{\arraystretch}{2}
\begin{tabular}{|c|c|c|c|}
\multicolumn{4}{|c|}{\emph{Union2.1+H(z)+AP}}\\
\hline
 & $x(a_{0})$ & $y(a_{0})$ & $\lambda(a_{0})$ \\
\hline
model 1a & $-0.1862^{+0.6136}_{-0.6186}$ & $0.8826^{+0.3068}_{-0.3093}$ & $-2.0244^{+0.2280}_{-0.2303}$ \\
model 1b & $-0.1733^{+1.2743}_{-1.2591}$ & $0.8891^{+0.6371}_{-0.6295}$ & $-2.0311^{+0.6460}_{-0.6493}$\\
model 2 & $-0.1741^{+0.3314}_{-0.3286}$ & $0.8886^{+0.1657}_{-0.1643}$ & $-2.0230^{+0.1099}_{-0.1075}$\\
\hline\hline
  \multicolumn{4}{|c|}{\emph{Union2.1+H(z)+AP+BAO}}\\
  \hline
& $x(a_{0})$ & $y(a_{0})$ & $\lambda(a_{0})$ \\
\hline
model 1a & $-0.1813^{+0.4909}_{-0.4761}$ & $0.8851^{+0.2454}_{-0.2381}$ & $-2.0455^{+0.1777}_{-0.1900}$\\
model 1b & $-0.1826^{+2.2818}_{-1.3340}$ & $0.8844^{+1.1406}_{-0.6670}$ & $-2.0175^{+1.0997}_{-0.6620}$\\
model 2 & $-0.1860^{+0.2126}_{-0.2023}$ & $0.8827^{+0.1063}_{-0.1012}$ & $-2.0312^{+0.0479}_{-0.0557}$\\
\end{tabular}
\end{ruledtabular}
\end{table}

With the present time value of the $y(a_{0})$ phase space variable at hand, one is able to calculate the scalar field potential function value at the present time, which is
\begin{equation}
V(\phi_{0}) = 3\phi_{0} H_{0}^{2} \big(y(a_{0})\big)^{2}\,.
\end{equation}
Additionally from the $\lambda(a_{0})$ we are able to calculate the first derivative of the scalar field potential function with respect to the scalar field as
\begin{equation}
V'(\phi_{0}) = - 3 H_{0}^{2} \big(y(a_{0})\big)^{2} \lambda(a_{0})\,.
\end{equation}

\begin{table}
\centering
\label{tab:4}
\caption{Values of the Brans-Dicke parameter $\om$ calculated for the mean of marginalized posterior PDF with 68\% confidence levels.}
\renewcommand{\arraystretch}{2}
\begin{tabular}{|c|c|c|}
\hline\hline
\multicolumn{2}{|c|}{\emph{Union2.1+H(z)+AP}}\\
\hline
 & \multicolumn{1}{c|}{$\om$} \\
\hline
model 1a  & $-0.7364^{+2.7579}_{-0.4284}$\\
model 1b  & $-2.4430^{+1.2743}_{-1.2591}$\\
model 2  & $-1.0780^{+0.3805}_{-0.2014}$ \\
\hline\hline
  \multicolumn{2}{|c|}{\emph{Union2.1+H(z)+AP+BAO}}\\
  \hline
 & \multicolumn{1}{c|}{$\om$} \\
\hline
model 1a & $-0.8606^{+0.8281}_{-0.1341}$ \\
model 1b & $-2.3837^{+0.4588}_{-4.5459}$\\
model 2  & $-1.1103^{+0.1872}_{-0.1729}$\\
\hline\hline
\end{tabular}
\end{table}

\begin{center}
\begin{figure}
\includegraphics[scale=0.30]{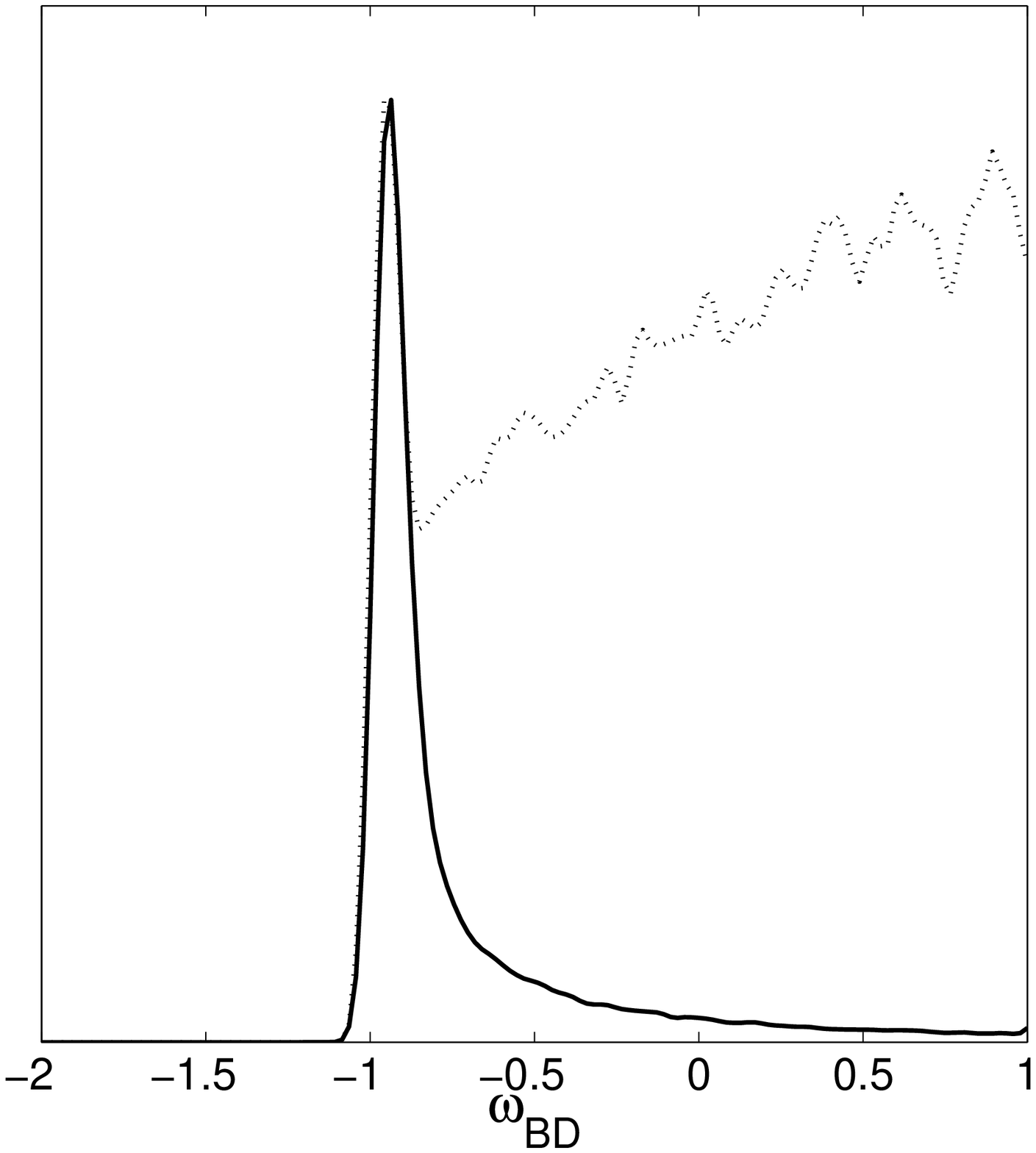}\\
\includegraphics[scale=0.30]{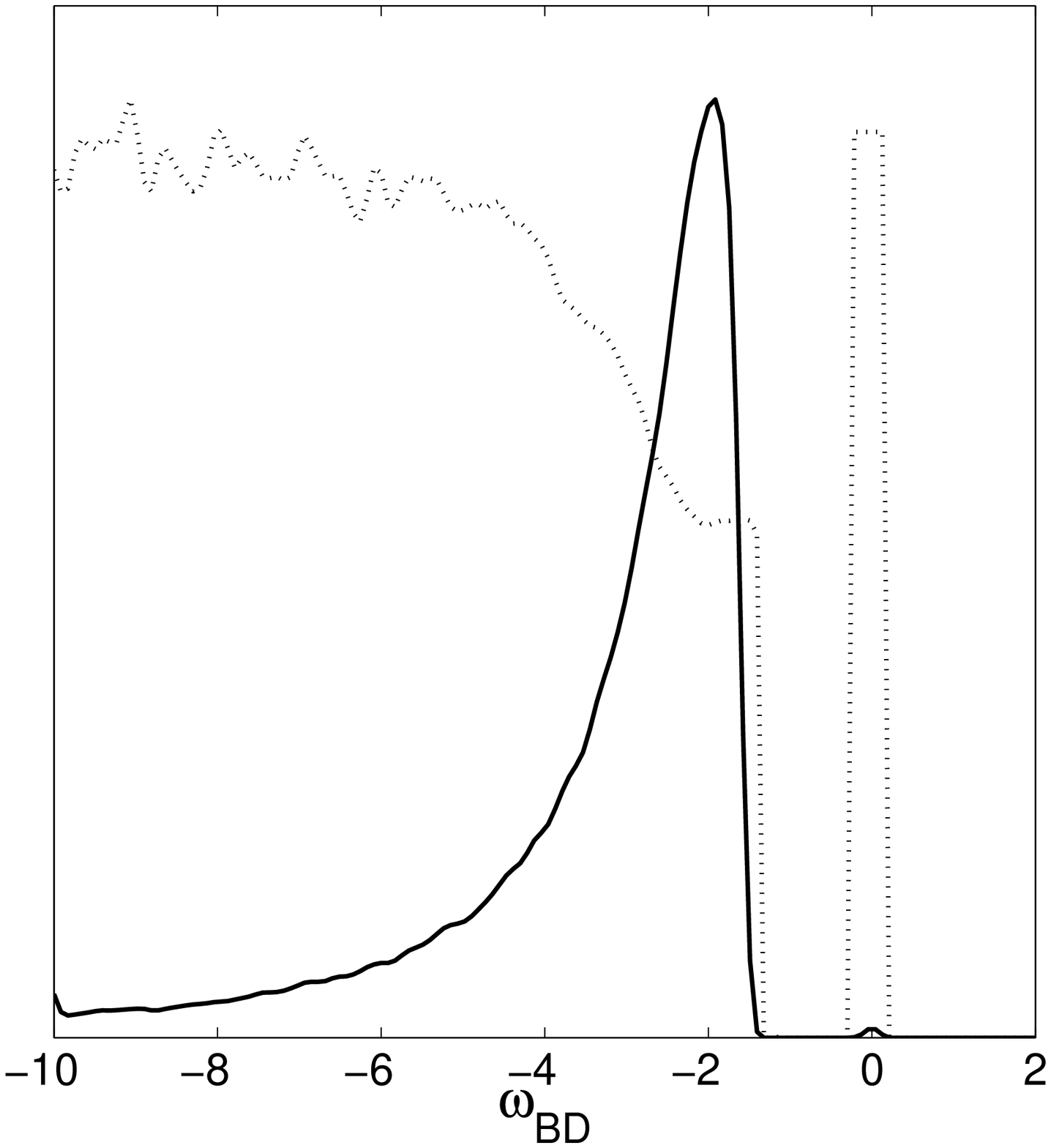}\\
\includegraphics[scale=0.30]{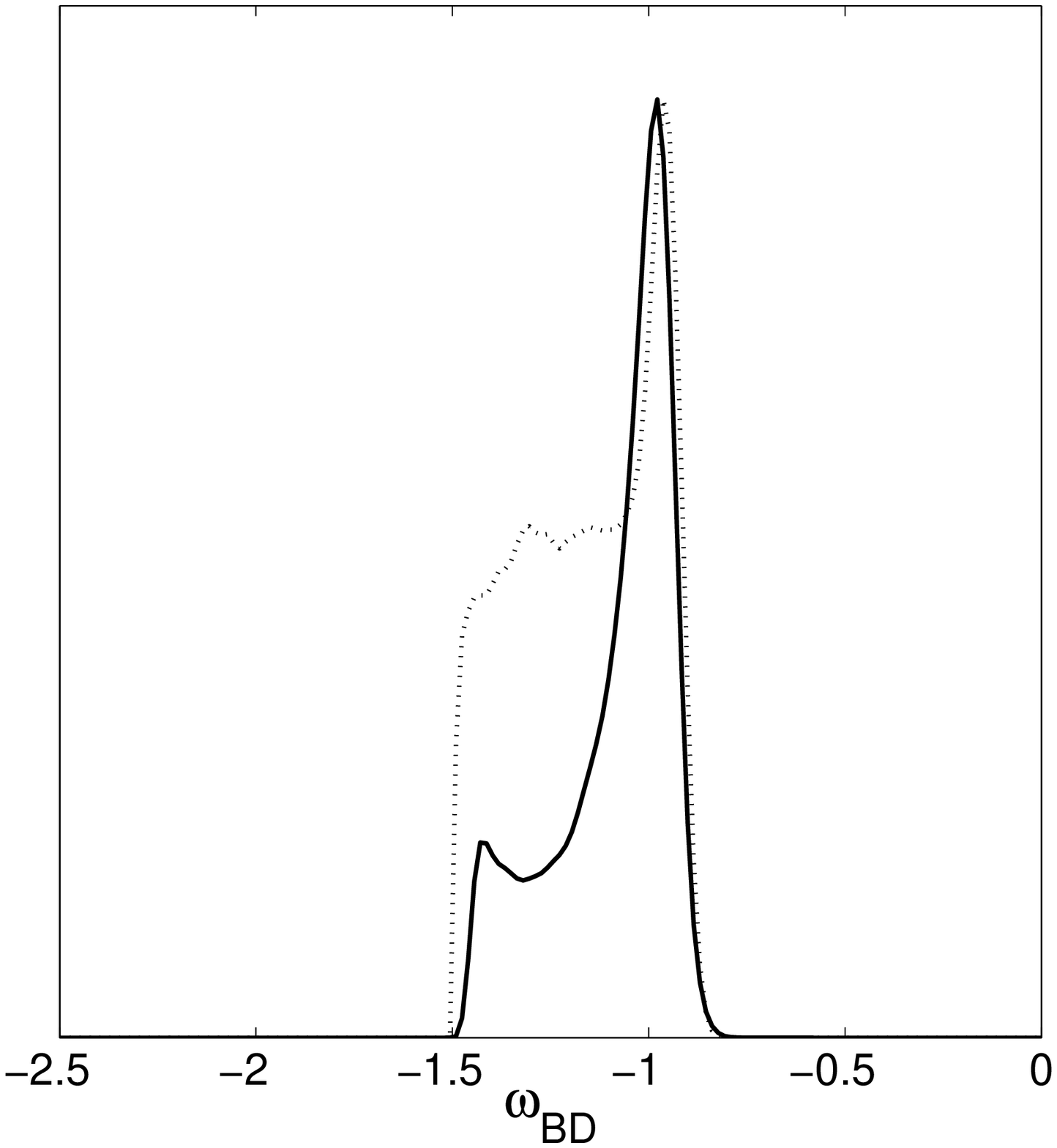}
\caption{The fully marginalized probabilities (solid lines) and mean likelihood (dotted lines) for the parameter of the Brans-Dicke theory $\om$ calculated for the models $1a$ (top) and $1b$ (middle) from Eq.~\eqref{eq:om_lin} and for the model $2$ (bottom) from Eq.~\eqref{eq:om_osc}.}
\label{fig:5}
\end{figure}
\end{center}

The most important parameter in the Brans-Dicke theory is the only free parameter of the theory, namely, $\om$ parameter.

From \eqref{eq:M_lin} and \eqref{eq:M_osc} we obtain that the $\delta$ parameter can be directly expressed as
\begin{equation}
\delta=\frac{16\,\Omega_{bm,0}}{3(\Omega_{bm,0}-\Omega_{M,0})}\,,
\end{equation}
where in our considerations $\Omega_{bm,0}$ is the fixed value and $\Omega_{M,0}$ was estimated from the astronomical observational data. Then from \eqref{eq:om_1} we have
\begin{equation}
\label{eq:om_lin}
\om = -\frac{3}{2} + \frac{6}{n(n-3)}\frac{\Omega_{bm,0}}{\Omega_{bm,0}-\Omega_{M,0}}
\end{equation}
and from \eqref{eq:om_2} we obtain
\begin{equation}
\label{eq:om_osc}
\om = -\frac{3}{2}-\frac{24}{9+4n^{2}}\frac{\Omega_{bm,0}}{\Omega_{bm,0}-\Omega_{M,0}}\,.
\end{equation}
Note, that there is only one possibility to obtain $\om\gg1$, namely when in \eqref{eq:om_lin} the estimated value of $n$ parameter is $n\approx0$ (or equivalently $n\approx3$). 

In Table \ref{tab:4} we gathered the values of the Brans-Dicke parameter $\om$ calculated for the mean of marginalized posterior PDF with 68\% confidence level for the parameters of the models while on figure \ref{fig:5} we present the fully marginalized probabilities and mean likelihood  for the $\om$ parameter of the Brans-Dicke theory. In the first case for the model 1a one can notice a clear cutoff at $\om=-3/2$ as the value leading to pathologies in the model \cite{Dabrowski:2005yn}.

Finally we can calculate the mass of the Brans-Dicke scalar field. In the Jordan frame we have \cite{Faraoni:2009km}
\begin{equation}
m^{2} = \frac{2}{3+2\om}\big(\phi V''(\phi)-V'(\phi)\big)\,,
\end{equation} 
which, transformed into the investigated phase space variables, is
\begin{equation}
m^{2} = \frac{6}{3+2\om} H^{2} y^{2}\lambda(1+\lambda\,\Gamma(\lambda))\,.
\end{equation}
The mass of the Brans-Dicke scalar field is dynamical quantity and changes during the evolution of universe as well it depends on the form of the scalar field potential function. Using linearized solution to the dynamics one can calculate not only its asymptotic value at the de Sitter state but also its present value.

From the third Eq.~\eqref{eq:dynsys_c} of the dynamical system \eqref{eq:dynsys} describing the evolution of models we obtain that at the present epoch
\begin{equation}
a \frac{\ud \lambda}{\ud a}\Big|_{0} = x(a_{0}) \lambda(a_{0})\Big(1-\lambda(a_{0})\big(\Gamma(\lambda(a_{0}))-1\big)\Big)\,,
\end{equation}
and using the linearized solutions \eqref{eq:sol_lin} and \eqref{eq:sol_osc} one can calculate the quantity on the left-hand side of the equation. Then we obtain the present value of the $\Gamma(\lambda(a_{0}))$ function which depends on the second derivative of the scalar field potential function.

Finally we can express the mass of the Brans-Dicke scalar field at the present epoch as
\begin{equation}
\begin{split}
m^{2}\big|_{0} = &\frac{6}{3+2\om} H^{2}(a_{0}) y^{2}(a_{0}) \times \\
&\times\Big(\big(2+\lambda(a_{0})\big)\lambda(a_{0}) - \frac{1}{x(a_{0})}a_{0} \frac{\ud \lambda}{\ud a}\Big|_{0}\Big)
\end{split}
\end{equation}
where from \eqref{eq:sol_lin_c} and \eqref{eq:l_lin} we have
\begin{equation}
a_{0}\frac{\ud\lambda}{\ud a}\Big|_{0} = \frac{3}{2}\Omega_{bm,0} + \frac{3}{8}\frac{n}{n+1}\Omega_{n,0}-\frac{3}{8} \frac{n-3}{n-4}\Omega_{3n,0}\,,
\end{equation}
while from \eqref{eq:sol_osc_c} and \eqref{eq:l_osc} we have
\begin{equation}
\begin{split}
a_{0}\frac{\ud\lambda}{\ud a}\Big|_{0} = &\frac{3}{2}\Omega_{bm,0} + \frac{2}{4n^{2}+25}
\frac{\Omega_{bm,0}}{\Omega_{bm,0}-\Omega_{M,0}}\times\\
&\times\Big(-(4n^{2}+15)\Omega_{cos,0}+4n\,\Omega_{sin,0}\Big)\,.
\end{split}
\end{equation}

For a linear approach to the de Sitter state (model 1a) we obtain the mass of the Brans-Dicke scalar field
\begin{equation}
m^{2}\big|_{0} = 2.3604_{-4.6626}^{+53.6042}\, H_{0}^{2}\,,
\end{equation}
while for a oscillatory approach to the de Sitter state (model 2) we have the following mass at the present epoch
\begin{equation}
m^{2}\big|_{0} = 2.8062_{-0.8873}^{+3.3998}\, H_{0}^{2}\,.
\end{equation}
In the model 1b the de Sitter state is represented by a saddle-type critical point and we obtain
\begin{equation}
m^{2}\big|_{0} = -1.2341_{-33.7046}^{+1.8205}\, H_{0}^{2}\,.
\end{equation}
In two first cases we obtain the mass of the Brans-Dicke scalar field as
\begin{equation}
m|_{0}\sim H_{0}
\end{equation}
which is consistent with an upper bound on the mass of a ultralight pseudo Nambu-Goldstone bosons considered in a cosmological background \cite{Frieman:1995pm,Amendola:2005ad}. In the model 1b where the de Sitter state is a transient state (represented by a saddle-type critical point) the mass of the Brans-Dicke scalar field is of a tachyonic type.

\section{Conclusions}

As we mentioned in the Introduction, the Cassini spacecraft mission in the parametrized post-Newtonian (PPN) formalism gave the most stringent experimental limit $\om>40000$ on the value of 
the Brans-Dicke parameter \cite{Bertotti:2003rm}. This was obtained in the solar system test for
spherically symmetric solutions. The cosmological constraints on the Brans-Dicke parameter 
$\om$ concern different spatial and temporal scales and the cosmography now plays the role of the PPN formalism. In order to obtain the Hubble functions we did not assume any specific form of the potential function for the Brans-Dicke scalar field. The chameleon mechanism \cite{Khoury:2003aq, Khoury:2003rn, Gubser:2004uf, Brax:2004qh, Mota:2006ed, Mota:2006fz} leads to modifications in the effective potential function, i.e., the effective mass of the scalar field, which depends on the local matter density. In regions of low-mass density like on the cosmological scales, the scalar field is light, while in regions of high density in the solar system, it acquires a large mass, making its effects unobservable. However, the chameleon mechanism is not a generic feature for arbitrary scalar field potential functions. The question whether this mechanism arises for all possible potential functions under considerations remains open.

From a theoretical point of view there are two special values of the Brans-Dicke parameter, namely $\om=0$ and $\om=-1$. 

In the metric formulation of $f(R)$ theory of gravity the action integral,
\begin{equation}
S=\int\ud^{4}x\sqrt{-g}\,f(R)\,,
\end{equation}
can be rewritten in the following form \cite{Capozziello:2011et}
\begin{equation}
S=\int\ud^{4}x\sqrt{-g}\big(\phi R-2V(\phi)\big)\,,
\end{equation}
which is equivalent to the Brans-Dicke theory with $\om=0$.

From the other hand, the Lagrangian density of the low-energy limit of the bosonic string theory
\cite{Green:book,Fradkin:1985ys,Tseytlin:1988rr} can be presented in the following form,
\begin{equation}
\mathcal{L} = e^{-2\Phi}\big(R + 4\nabla^{\alpha}\Phi\,\nabla_{\alpha}\Phi - \Lambda\big)\,,
\end{equation}
where $\Phi$ is the dilaton field. Making the substitution $\phi=e^{-2\Phi}$, one obtains the 
Brans-Dicke theory with $\om=-1$ and $V(\phi)=\Lambda\phi$. Neglecting the matter, the two 
theories are identical, but they differ in their couplings of the scalar field to the other 
matter \cite{Kolitch:1994qa}.

In this paper we obtained cosmological constraints on the models resulting from dynamical analysis of the Brans-Dicke theory. We have shown that for an arbitrary potential function of the Brans-Dicke scalar field, there exists the de Sitter state and that the dynamical behavior in its vicinity crucially depends on the value of the first and second derivative of the scalar field potential function at the de Sitter state as well as on the value of the Brans-Dicke parameter. We found the following mean values of the parameter of the theory: for a linear approach to the de Sitter state $\om=-0.8606^{+0.8281}_{-0.1341}$, for an oscillatory approach to the de Sitter state we obtain $\om=-1.1103^{+0.1872}_{-0.1729}$ while for the transient de Sitter state represented by a saddle-type critical point we find $\om=-2.3837^{+0.4588}_{-4.5459}$. It is interesting that for the models under investigation, for an arbitrary scalar field potential function and excluding the model with $\om<-3/2$ as one leading to ghost behavior, we obtained a value of the $\om$ parameter close to the value needed to obtain correspondence with the low-energy limit of the bosonic string theory.

\begin{acknowledgments}
We are very grateful to Adam Krawiec for valuable suggestions and comments.

The research of O.H. was supported by the Polish Ministry of Science and Higher
Education through the project ``Iuventus Plus'' (Contract No.~0131/H03/2010/70) and by the
National Science Centre through the postdoctoral internship award 
(Decision No.~DEC-2012/04/S/ST9/00020). M.S. was supported by the National Science Centre through the OPUS 5 funding scheme (Decision No.~DEC-2013/09/B/ST2/03455), and M.K. was supported by the National Science Centre through the PRELUDIUM funding scheme (Decision No.~DEC-2012/05/N/ST9/03857). The use of the {\'S}wierk Computing Centre (CI{\'S}) computer cluster at the National Centre for Nuclear Research is gratefully acknowledged.
\end{acknowledgments}

\newpage
\begin{widetext}
\appendix
\section{Linearized solutions in the vicinity of the de Sitter state}
\label{app_a}
Here we present the complete form of the linearized solutions of the system \eqref{eq:dynsys} in the vicinity of the de Sitter state.
In the case of purely real eigenvalues (model 1a and model 1b) using substitution 
\begin{equation}
\frac{\delta}{3+2\om}=\frac{4}{9}n(n-3)\,,
\end{equation}
the eigenvalues of the linearization matrix take the following form,
\begin{equation}
l_{1}=-3\,, \quad l_{2}=-n\,, \quad l_{3}=-3+n\,,
\end{equation}
and the linearized solutions are
\begin{subequations}
\label{eq:sol_lin}
\begin{align}
x(a) & =  \frac{4}{\delta}\big(\Delta x -2 \Delta y\big)\bigg(\frac{a}{a^{(i)}}\bigg)^{-3} 
+\frac{n}{3\delta(2n-3)}\Big(-4n\big(\Delta x-2\Delta y\big)+3\delta\Delta x-8(n-3)\Delta \lambda\Big)\bigg(\frac{a}{a^{(i)}}\bigg)^{-n} +\nonumber \\ &
+\frac{n-3}{3\delta(2n-3)}\Big(4(n-3)\big(\Delta x -2\Delta y\big)+3\delta \Delta x +8n\Delta\lambda\Big)\bigg(\frac{a}{a^{(i)}}\bigg)^{-3+n}\,,\\
y(a) & = 1+\frac{1}{2}\left(\frac{4}{\delta}-1\right)\big(\Delta x-2\Delta y\big)\bigg(\frac{a}{a^{(i)}}\bigg)^{-3} + 
\frac{n}{6\delta(2n-3)}\Big(-4n\big(\Delta x-2\Delta y\big)+3\delta\Delta x-8(n-3)\Delta \lambda\Big)\bigg(\frac{a}{a^{(i)}}\bigg)^{-n} +\nonumber \\ & 
+\frac{n-3}{6\delta(2n-3)}\Big(4(n-3)\big(\Delta x -2\Delta y\big)+3\delta \Delta x +8n\Delta\lambda\Big)\bigg(\frac{a}{a^{(i)}}\bigg)^{-3+n}\,,\\
\label{eq:sol_lin_c}
\lambda(a) & =  -2-\frac{1}{2}\big(\Delta x-2\Delta y\big)\bigg(\frac{a}{a^{(i)}}\bigg)^{-3} 
+\frac{1}{8\delta(2n-3)}\Big(-4n\big(\Delta x-2\Delta y\big)+3\delta\Delta x-8(n-3)\Delta \lambda\Big)\bigg(\frac{a}{a^{(i)}}\bigg)^{-n} +\nonumber \\ & 
+\frac{1}{8\delta(2n-3)}\Big(4(n-3)\big(\Delta x -2\Delta y\big)+3\delta \Delta x +8n\Delta\lambda\Big)\bigg(\frac{a}{a^{(i)}}\bigg)^{-3+n}\,,
\end{align}
\end{subequations}
where $\Delta x = x^{(i)}$, $\Delta y = y^{(i)}-1$, and $\Delta\lambda= \lambda^{(i)}+2$ are the 
initial conditions.

In the case of the eigenvalues with nonzero imaginary part (model 2) using the substitution
\begin{equation}
\frac{\delta}{3+2\om}=-\frac{1}{9}(9+4n^{2})\,,
\end{equation}
the eigenvalues of the linearization matrix take the following form,
\begin{equation}
l_{1}=-3\,, \quad l_{2}=-\frac{3}{2}-\mathbbmtt{i} n, \quad l_{3}=-\frac{3}{2}+\mathbbmtt{i} n\,,
\end{equation}
and the linearized solutions are
\begin{subequations}
\label{eq:sol_osc}
\begin{align}
x(a) & = \frac{4}{\delta}\big(\Delta x -2 \Delta y\big)\bigg(\frac{a}{a^{(i)}}\bigg)^{-3} 
+ \left(-\frac{4}{\delta}\big(\Delta x -2 \Delta y\big) + \Delta x\right)\bigg(\frac{a}{a^{(i)}}\bigg)^{-3/2}\cos{\left(n\,\ln{\bigg(\frac{a}{a^{(i)}}}\bigg)\right)}+\nonumber\\ 
& + \frac{1}{6n}\left(-\frac{2}{\delta}(4n^{2}-9)\big(\Delta x -2 \Delta y\big) - 9 \Delta x -\frac{4}{\delta}(4n^{2}+9)\Delta\lambda\right)\bigg(\frac{a}{a^{(i)}}\bigg)^{-3/2}\sin{\left(n\,\ln{\bigg(\frac{a}{a^{(i)}}}\bigg)\right)}\,,\\
y(a) & = 1 + \frac{1}{2}\left(\frac{4}{\delta}-1\right)\big(\Delta x -2 \Delta y\big)\bigg(\frac{a}{a^{(i)}}\bigg)^{-3} 
+ \frac{1}{2}\left(-\frac{4}{\delta}\big(\Delta x -2 \Delta y\big) + \Delta x\right)\bigg(\frac{a}{a^{(i)}}\bigg)^{-3/2}\cos{\left(n\,\ln{\bigg(\frac{a}{a^{(i)}}}\bigg)\right)}+\nonumber\\ 
& + \frac{1}{12n}\left(-\frac{2}{\delta}(4n^{2}-9)\big(\Delta x -2 \Delta y\big) - 9 \Delta x -\frac{4}{\delta}(4n^{2}+9)\Delta\lambda\right)\bigg(\frac{a}{a^{(i)}}\bigg)^{-3/2}\sin{\left(n\,\ln{\bigg(\frac{a}{a^{(i)}}}\bigg)\right)} \,,\\
\label{eq:sol_osc_c}
\lambda(a) & = -2-\frac{1}{2}\big(\Delta x -2 \Delta y\big)\bigg(\frac{a}{a^{(i)}}\bigg)^{-3}
+\left(\frac{1}{2}\big(\Delta x -2 \Delta y\big)+\Delta\lambda\right)\bigg(\frac{a}{a^{(i)}}\bigg)^{-3/2}\cos{\left(n\,\ln{\bigg(\frac{a}{a^{(i)}}}\bigg)\right)}+\nonumber\\ 
& +\frac{3}{2n}\left(-\frac{1}{2}\big(\Delta x -2\Delta y\big) + \frac{\delta}{4}\Delta x + \Delta\lambda\right)\bigg(\frac{a}{a^{(i)}}\bigg)^{-3/2}\sin{\left(n\,\ln{\bigg(\frac{a}{a^{(i)}}}\bigg)\right)}\,,
\end{align}
\end{subequations}
where again $\Delta x = x^{(i)}$, $\Delta y = y^{(i)}-1$, and $\Delta\lambda= \lambda^{(i)}+2$ are the 
initial conditions.
\end{widetext}

\bibliographystyle{apsrev4-1}
\bibliography{../moje,../darkenergy,../quintessence,../quartessence,../astro,../dynamics,../standard,../inflation,../bd_theory,../strings}

\end{document}